\documentclass[aps,prd,twocolumn,superscriptaddress,preprintnumbers,nofootinbib]{revtex4-2}
\usepackage[utf8]{inputenc}
\usepackage{amsmath,amssymb,amsfonts}
\usepackage{graphicx}
\usepackage{xcolor}
\definecolor{revcolor}{rgb}{0.70,0,0}

\newcommand{\Tr}{\operatorname{Tr}}
\usepackage[colorlinks=true, allcolors=blue]{hyperref}
\usepackage{orcidlink}

\makeatletter  
\renewcommand\onecolumngrid{
\do@columngrid{one}{\@ne}%
\def\set@footnotewidth{\onecolumngrid}
\def\footnoterule{\kern-6pt\hrule width 1.5in\kern6pt}%
}
\makeatother    
\begin{document}
\begin{figure}
\vskip -1.cm
\leftline{\includegraphics[width=0.15\textwidth]{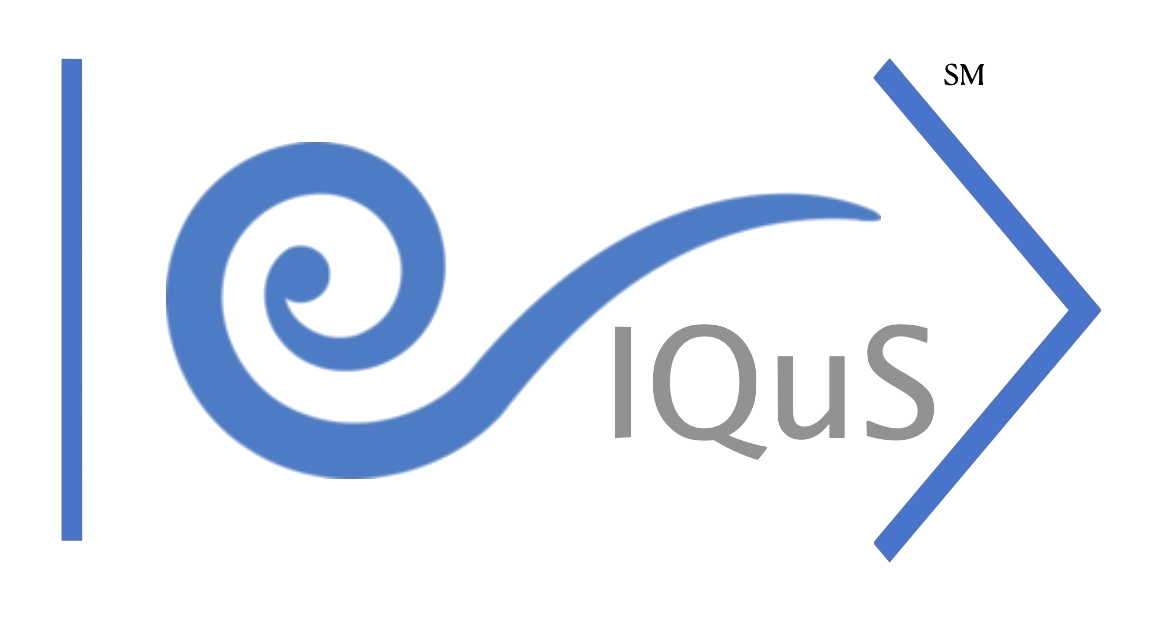}}
\end{figure}
\preprint{IQuS@UW-21-125, NT@UW-26-14}

\author{Sebastian Grieninger\,\orcidlink{0000-0002-9523-5819}}
\email{segrie@uw.edu}
\affiliation{InQubator for Quantum Simulation (IQuS), Department of Physics, University of Washington, Seattle, WA 98195, USA}
\title{Nonlocal Nonstabilizerness from Holographic Schwinger Pair Production}
\date{\today}

\begin{abstract}
We analyze the emergence of nonlocal magic in 
Schwinger pair creation in strong non-Abelian (chromo)electric 
fields using holography. The produced quark--antiquark pair is 
entangled into a color singlet, yet accelerates into causally 
disconnected Rindler wedges. Using the Casini--Huerta--Myers 
conformal mapping and the probe-brane framework, we compute the 
refined R\'enyi entropy and its derivative, which 
captures the antiflatness of the entanglement spectrum for a spherical bipartition. 
We find that for boundary spacetime dimension $d>2$, the 
entanglement spectrum is non-flat, implying the dynamical 
generation of nonlocal magic in the pair creation process. Interestingly, 
the nonlocal magic in the holographic dual can be obtained from the free energy of the probe action. 
\end{abstract}

\maketitle
\section{Introduction}
The emergence of nonlocal quantum correlations in nonequilibrium processes is a central question in quantum many body systems~\cite{Robin:2026lqp}. While entanglement entropy quantifies the amount of correlations generated, it does not capture their internal structure.

In this work, we focus on nonlocal magic: the part of non-stabilizerness across a bipartition that cannot be removed by local unitary transformations. We use the term
``magic'' in this resource-theoretic sense, rather than as circuit complexity. While magic is closely related to the difficulty of classical simulation in stabilizer-based descriptions, it is not identical to
algorithmic quantum complexity and should not be treated as a universal
measure of computational hardness.

Recent developments in quantum information science have emphasized that entanglement alone does not fully characterize the structure of quantum many-body states~\cite{Haferkamp:2021uxo, Chitambar:2018rnj,Brown:2017jil,Eisert:2008ur,Leone:2021rzd,Robin:2020aeh,Haug:2023hcs,Tarabunga:2023hau,Hengstenberg:2023ryt,Haug:2024ptu,Emerson:2013zse,Howard:2017maw,Hamaguchi:2023zpb,Tirrito:2023fnw,Chernyshev:2024pqy,Cao:2024nrx,Robin:2024oqc,brokemeier2025quantum,Robin:2025ymq,Jiang:2025wpj,White:2020zoz,Cao:2026uoq,Hou:2025bau,Ebner:2025pdm,Grieninger:2026bdq,Xu:2026ibi,Joshi:2026hfe,Cao:2026qky,Halimeh:2025vvp,Iannotti:2026jvd,Collura:2026hlz}. While entanglement entropy captures the total amount of quantum correlations, it is insensitive to how these correlations are organized within the Hilbert space. Measures of quantum complexity, such as nonlocal magic~\cite{Cao:2023mzo,Cao:2024nrx}, provide complementary information by probing the microscopic structure underlying these correlations.

In particular, nonlocal magic is sensitive to fluctuations and variance of the entanglement spectrum~\cite{Cao:2024nrx,Tirrito:2023fnw}, revealing features that remain invisible to the entanglement entropy~\cite{Grieninger:2026bdq}. This distinction has proven important in a range of physical settings, where states with similar entanglement can exhibit vastly different internal structures and dynamical properties~\cite{li2008entanglement,Yang:2017myj}. For example, nonlocal magic can detect nontrivial organization in many-body wavefunctions even when conventional observables or classical order parameters appear simple, and can expose transient regimes of high complexity during dynamical evolution that act as barriers to efficient simulation~\cite{Haferkamp:2021uxo,Ebner:2025pdm}.

From this perspective, complexity measures provide a refined diagnostic of quantum correlations, capturing how information is distributed and processed across a system (see also holographic magic-harvesting approaches
in Refs~\cite{Yang:2025zrl,Zhang:2026jll}). This additional sensitivity makes them particularly well suited for probing far-from-equilibrium dynamics, strongly coupled systems, and emergent phenomena in quantum field theory.

Schwinger pair creation~\cite{PhysRev.82.664} in strong chromoelectric fields offers a natural setting to study this question. In this process, an external field pulls quark--antiquark pairs out of the vacuum, similar to the strong color fields generated in the early stages of high-energy heavy-ion collisions.

At strong coupling, the produced pair is described holographically by an open string whose endpoints accelerate apart, providing a controlled framework in which both the tunneling process and the subsequent real-time dynamics can be analyzed. The Schwinger pair provides a concrete realization of EPR 
entanglement in gauge theory~\cite{Jensen:2013ora,Sonner:2013mba,
Jensen:2014lua}. The produced quark and antiquark are entangled into a color singlet yet accelerate into causally disconnected Rindler wedges. Connected correlators between the pair are nonzero but cannot arise from interactions after nucleation, identifying the pair's entanglement as EPR-type. Hence, it's natural to ask what quantum correlations the Schwinger pair carries beyond its total entanglement.

This problem is closely related to string formation and breaking in confining gauge theories. In such systems, (color) charges are connected by flux tubes whose dynamics govern hadronization and particle production. Recent studies of lattice models in one and two spatial dimensions, respectively, using tensor-network and quantum simulation methods have shown that quantum complexity measures, including entanglement and magic, reveal nontrivial structure in the wavefunction during string formation and fragmentation~\cite{Buyens:2015tea,Grieninger:2025rdi,Florio:2025hoc, Florio:2023dke, Florio:2024aix, Grieninger:2026bdq,Barata:2025hgx,Artiaco:2025qqq,Verdel:2019chj,Verdel:2023mmp,Mallick:2024slg,Cochran:2024rwe,Gonzalez-Cuadra:2024xul, Borla:2025gfs,Cataldi:2025cyo,Xu:2025abo,DiMarcantonio:2025cmf,Ciavarella:2024fzw,Ciavarella:2024cyt,Crippa:2024hso,Liu:2024lut,De:2024smi,Surace:2024bht,Ciavarella:2024lsp,Alexandrou:2025vaj,Luo:2025qlg}. These results suggest that particle production processes are accompanied not only by the generation of entanglement, but also by a reorganization of quantum correlations.

In this work, we investigate the emergence of nonlocal quantum correlations in the pair creation process by analyzing the entanglement structure of a fixed real-space subsystem containing one member of the pair (i.e. a half-space bipartition). We find that the Lorentzian structure reshapes the entanglement spectrum and leads to nonlocal magic for $d>2$.

\section{Nonlocal Magic of the Schwinger pair}
\label{sec:schwinger}

\subsection{String worldsheet as an AdS$_2$ black hole}
\label{sec:ws_bh}

We consider a quark--antiquark pair produced by the Schwinger
mechanism in a large-$N_c$, large-$\lambda$ gauge theory with
a holographic dual. In this limit string breaking is suppressed,
and the pair remains connected by an open string. Fundamental
matter is introduced via a probe Dirichlet-brane at radial position
$z = z_M$, on which the quark has
mass~\cite{Grieninger:2023ehb,Grieninger:2023pyb,Xiao:2008nr,Semenoff:2011ng,Lewkowycz:2013laa,Chernicoff:2013iga,Jensen:2014bpa,Hubeny:2014zna,Ghodrati:2015rta,Semenoff:2018ffq}
\begin{equation}
M = \frac{\sqrt{\lambda}}{2\pi\, z_M}\,.
\label{eq:quark_mass}
\end{equation}
A constant background electric field $E$ on the brane induces
pair production. The holographic description of this process
is a string worldsheet in $\mathrm{AdS}_{d+1}$ governed by the
Nambu--Goto action~\cite{Semenoff:2011ng},
\begin{equation}
S_{\rm NG} = \frac{\sqrt{\lambda}}{4\pi}
\int d\tau \int d\sigma\;
\frac{1}{z^2}\bigl(\partial X_\mu\, \bar{\partial} X_\mu
+ \partial z\, \bar{\partial} z \bigr)
+ i \oint A\,,
\label{eq:NG}
\end{equation}
where $(\partial, \bar{\partial}) = (\partial_\sigma \pm i\,
\partial_\tau)$ and Virasoro constraints are imposed. The
boundary term $i\oint A$ captures the Lorentz force on the
quark endpoints.

The Euclidean instanton describing the tunnelling event has
the quark worldline tracing a circular path of cyclotron
radius~$R_{\rm cyc}$ on the Dirichlet-brane. The worldsheet solution 
satisfying the equations of motion and Virasoro constraints was 
obtained in Ref~\cite{Semenoff:2011ng}; extremizing the on-shell 
action over~$R_{\rm cyc}$ fixes the cyclotron radius through
\begin{equation}
R_{\rm cyc}^2 + z_M^2 = \frac{1}{a^2}\,,
\label{eq:cyclotron}
\end{equation}
where $a \equiv E/M$. The instanton exists for $E \leq E_c$, with 
critical field $E_c = M/z_M$ corresponding to $R_{\rm cyc} = 0$. Note that the pair appears spatially separated in real space at $t=0$, i.e. it appears nonlocally in the Lorentzian spacetime.

The induced metric on the worldsheet
instanton~\cite{Semenoff:2011ng} is
\begin{equation}
ds^2_W = \frac{1}{\sinh^2(2\pi n_{\rm inst}\,\sigma)}
\bigl(d\tau^2 + d\sigma^2\bigr)\,,
\label{eq:ws_Euclidean}
\end{equation}
where $n_{\rm inst}$ is the instanton winding number. A single pair 
corresponds to $n_{\rm inst} = 1$. 

Under analytic continuation to 
Lorentzian signature, $\tau \to it$, this becomes
\begin{equation}
ds^2_W = \frac{1}{\sinh^2(2\pi\sigma)}
\bigl(-dt^2 + d\sigma^2\bigr)\,.
\label{eq:ws_Lor}
\end{equation}
Introducing the coordinate
$w = \coth(2\pi\sigma)$, the metric takes the manifestly
static form
\begin{equation}
ds^2_W = -(w^2 - 1)\, dt^2
+ \frac{dw^2}{w^2 - 1}\,,
\label{eq:ws_bh}
\end{equation}
which is an $\mathrm{AdS}_2$ black hole with horizon at
$w = 1$.

The Hartle--Hawking construction patches the Euclidean
instanton to the Lorentzian solution at $t = 0$ hence producing the
maximally extended two-sided geometry, whose Penrose diagram
has two asymptotic boundaries (one near each string endpoint)
connected through the horizon~\cite{Jensen:2014bpa}. This
worldsheet wormhole is the holographic geometrization of EPR
entanglement between quark and antiquark: whereas entanglement
among $\mathcal{O}(N^2)$ degrees of freedom gives rise to
spacetime wormholes, the $\mathcal{O}(\sqrt{\lambda})$
entanglement carried by the colour-singlet pair is encoded on
the string worldsheet~\cite{Jensen:2014bpa,Maldacena:2013xja}.
Note that the Hartle--Hawking construction yields a thermofield-double state 
of the form
\begin{equation}
|\Psi\rangle \sim \sum_{k,c,\bar{c}} e^{-\pi E_k/a}\, 
\delta^{c\bar{c}}\, |k,c\rangle_q \otimes |k,\bar{c}\rangle_{\bar{q}}\,,
\label{eq:TFD}
\end{equation}
where $|k,c\rangle_q$ are eigenstates of the Rindler Hamiltonian 
with energy~$E_k$ and color index~$c$, and the inverse temperature 
is $\beta_U = 2\pi/a$. We emphasize, however, that the entanglement 
we analyze is not the entanglement of the color singlet structure 
eq.~\eqref{eq:TFD}, but rather the entanglement of a spherical 
spatial region~$A$ containing one member of the pair with its 
complement~$\bar{A}$ which is equivalent to the quark--antiquark (bipartite) entanglement.\footnote{The position dependence of the entanglement entropy of the accelerated pair was studied in Ref.~\cite{Argandona:2025msq}. The symmetric cut used in this work is the unique bipartition (which coincides with the quark--antiquark bipartition) and the reduced state is thermal on $S^1\times\mathbb{H}^{d-1}$.} The reduced density matrix~$\rho_A$ therefore 
encodes the interplay of the pair's color-singlet structure with the 
full QFT vacuum.

\subsection{Casini--Huerta--Myers (CHM) map and topological black holes}
\label{sec:CHM}

To compute R\'enyi entropies for a spherical entangling region
of radius~$R_A$ we employ the conformal mapping of Casini,
Huerta, and Myers~\cite{Casini:2011kv}. The
Euclidean $\mathrm{AdS}_{d+1}$ in Poincar\'e coordinates,
\begin{equation}
ds^2 = \frac{L^2}{r^2}\, dr^2
+ \frac{r^2}{L^2}\!\left(dt_E^2 + d\rho^2
+ \rho^2\, d\Omega_{d-2}^2\right),
\label{eq:Poincare}
\end{equation}
can be mapped to~\cite{Emparan:1999gf,Casini:2011kv}
\begin{equation}
ds^2 = L^2\!\left[\frac{d\zeta^2}{f(\zeta)}
+ f(\zeta)\, d\tau^2
+ \zeta^2\, du^2
+ \zeta^2 \sinh^2\!u\; d\Omega_{d-2}^2\right],
\label{eq:hyp_metric}
\end{equation}
with $f(\zeta) = \zeta^2 - 1$ (see appendix \ref{app:CHM_full} for details).

This coordinate system has two remarkable properties. First,
in the field theory, the transformation maps the causal
development of the spherical region~$A$ onto
$S^1_\tau \times \mathbb{H}^{d-1}$, where $\mathbb{H}^{d-1}$
is the hyperbolic plane parametrized by
$(u, \Omega_{d-2})$~\cite{Casini:2011kv}. The reduced density
matrix $\rho_A$ thereby becomes a thermal density matrix on
$\mathbb{H}^{d-1}$ at inverse temperature $\beta = 2\pi$.
Second, the metric~\eqref{eq:hyp_metric} has a Euclidean
horizon at $\zeta_h = 1$, which corresponds in Poincar\'e coordinates
to $r^2 + \rho^2 = R_A^2$. This is precisely the Ryu--Takayanagi
surface of the spherical
region~$A$~\cite{Ryu:2006bv,Ryu:2006ef}. The entanglement
entropy maps to the Bekenstein--Hawking entropy of this
horizon. Since the transformation is a coordinate change, the
geometry remains globally $\mathrm{AdS}_{d+1}$ and the horizon
is of Rindler
type~\cite{Emparan:1999gf,Casini:2011kv}.

The $n$-th R\'enyi entropy is obtained by extending the
periodicity to $\tau \sim \tau + 2\pi n$, which in the bulk
corresponds to replacing $f(\zeta)$ by the topological black
hole~\cite{Emparan:1999gf,Chalabi:2020tlw}
\begin{align}
f_n(\zeta) &= \zeta^2 - 1
- \frac{\zeta_h^{\,d} - \zeta_h^{\,d-2}}{\zeta^{d-2}}\,,
\label{eq:fn}\\[4pt]
\zeta_h(n) &= \frac{\sqrt{1 + n^2\, d(d{-}2)} + 1}{n\, d}\,,
\label{eq:zetah}
\end{align}
where smoothness of the Euclidean geometry with period
$2\pi n$ fixes $\zeta_h(n)$. At $n = 1$ one recovers
$\zeta_h = 1$ and $f_1 = \zeta^2 - 1$. Here $d$ denotes the
spacetime dimension of the boundary gauge theory ($d = 4$ for
$\mathcal{N} = 4$ SYM dual to $\mathrm{AdS}_5 \times S^5$),
so the topological black hole lives in the $(d+1)$-dimensional
bulk. Although the string worldsheet is two-dimensional, it is
embedded in this $(d+1)$-dimensional bulk and inherits its
$n$-dependence from the ambient geometry.

The relation between this two-dimensional worldsheet and the
$(d{+}1)$-dimensional bulk is the construction used by~\cite{Lewkowycz:2013laa}
for the entanglement entropy of a quark. The accelerated pair, equivalently a static quark at the center of
the sphere $A$ after the conformal map of Appendix~\ref{app:CHM_full}, maps
under the CHM map to a Polyakov loop at the center
$u=0$ of $\mathbb{H}^{d-1}$, wrapping $\tau$. Its holographic image is the probe
string at $u=0$ running from the brane to the horizon $\zeta_h$, and at $n=1$ the
induced metric of this string (Appendix~\ref{app:renyi_full}) reduces to the
AdS$_2$ black hole of Sec.~\ref{sec:ws_bh}, whose horizon is the RT surface
$\zeta_h$.

The same accelerated worldsheet underlies both the EPR pair of
Sec.~\ref{sec:ws_bh} and the pair produced here by tunneling; the instanton
fixes only the separation $R_{\rm cyc}$~\eqref{eq:cyclotron}, which does not
enter the $n$-dependence and hence, as we will see, affects neither $S_{EE}$
nor $C_E$. The quark endpoint is fundamental matter on the brane, so the
worldsheet ends on the brane in the UV and caps off on the horizon $\zeta_h$
(the RT surface) in the IR, at fixed $u=0$.

\subsection{R\'enyi entropies from the probe action}
\label{sec:renyi_probe}

We now connect the topological black hole
backgrounds~\eqref{eq:fn}--\eqref{eq:zetah} to R\'enyi
entropies, following the approach of Lewkowycz and
Maldacena~\cite{Lewkowycz:2013nqa} in probe brane language~\cite{Karch:2014ufa,Chalabi:2020tlw}. This is motivated by the quantum
information-theoretic connection between the refined R\'enyi entropy 
and nonlocal magic as follows (see~\cite{Cao:2024nrx} and appendices for details). The capacity of entanglement, 
$C_E = -\partial_n\widetilde{S}_n|_{n=1}$, equals the variance of 
the modular Hamiltonian spectrum, 
$C_E = \mathrm{Var}_\rho(-\log\rho_A)$. By Lemma~1 
of~\cite{Cao:2024nrx}, $C_E$ vanishes if and only if the 
entanglement spectrum is flat, which is equivalent to vanishing 
nonlocal magic (lower bounded by antiflatness):
\begin{equation}
C_E(\rho_A) = 0 
\;\Longleftrightarrow\; 
\mathcal{M}^{(\mathrm{NL})}(\psi_{AB}) = 0\,.
\label{eq:faithfulness}
\end{equation}
Thus $C_E > 0$ is both necessary and sufficient for the presence of 
nonlocal magic. This relation is purely information-theoretic.

In QFT, the $n$-th R\'enyi entropy
$S_A^{(n)} = \frac{1}{1-n}\log\Tr\rho_A^n$ may be computed
via the replica trick: $\Tr\rho_A^n$ equals the partition
function on a manifold $\mathcal{M}_n$ with a conical deficit
$2\pi/n$ at the entangling surface. Defining the generating
functional $W[\mathcal{M}_n] = -\log Z[\mathcal{M}_n]$, the
entanglement entropy
is~\cite{Callan:1994py,Holzhey:1994we,Calabrese:2004eu}
\begin{equation}
S_A = \lim_{n\to 1}(\partial_n - 1)\, W[\mathcal{M}_n]\,.
\label{eq:replica}
\end{equation}
In holography, the $n$-th R\'enyi
entropy follows from
$S_A^{(n)} = \frac{n}{n-1}\bigl[\widehat{I}(n)
- \widehat{I}(1)\bigr]$, $\hat I(n)$ is the on-shell gravitational probe action $\hat I(n)$~\cite{Dong:2016fnf}
\begin{equation}
\widetilde{S}_n = n^2\, \partial_n \widehat{I}(n)\,.
\label{eq:refined_renyi}
\end{equation}
 The derivative at $n = 1$ yields
the capacity of entanglement (see appendix \ref{app:CHM_full} for more details),
\begin{equation}
C_E = -\partial_n \widetilde{S}_n \big|_{n=1}\,,
\label{eq:capacity}
\end{equation}
which through~\eqref{eq:faithfulness}
provides a direct diagnostic for nonlocal magic in the state
of the produced pair. We emphasize that the vanishing of
boundary terms at all orders in the $n$-expansion ensures that
eq.~\eqref{eq:capacity} is obtained from
eq.~\eqref{eq:refined_renyi} without additional contributions.

\subsection{Nonlocal magic of the Schwinger pair}
\label{sec:NL_magic}

We now evaluate the refined R\'enyi entropy
$\partial_n\widetilde{S}_n|_{n=1}$ for the probe string.\footnote{The dominant vacuum contribution to the entanglement entropy
of the spherical region is of order $N_c^2$ and is independent
of the produced pair. Our calculation isolates the
$\mathcal{O}(\sqrt{\lambda})$ probe-string contribution
associated with the heavy quark--antiquark pair. All entropies
and R\'enyi derivatives below are therefore excess
contributions relative to the vacuum, induced by the probe.}

The on-shell Nambu--Goto action in the $n$-th
topological black hole background is (up to a constant; see appendix \ref{app:renyi_full} for more details)
\begin{equation}
\widehat{I}(n) = -\sqrt{\lambda}\;\zeta_h(n)\,.
\label{eq:In}
\end{equation}
The refined R\'enyi entropy~\eqref{eq:refined_renyi} is
therefore
\begin{equation}
\widetilde{S}_n = n^2\,\partial_n \widehat{I}(n)
= -\sqrt{\lambda}\; n^2\, \zeta_h'(n)\,.
\label{eq:Stilde_wh}
\end{equation}
At $n = 1$ this gives the entanglement entropy (in agreement with
Refs~\cite{Jensen:2013ora,Lewkowycz:2013laa,Kumar:2017vjv,Chalabi:2020tlw,Grieninger:2023ehb}),
\begin{equation}
S_{EE} = \widetilde{S}_1
= -\sqrt{\lambda}\;\zeta_h'(1)
= \frac{\sqrt{\lambda}}{d-1}\,.
\label{eq:SEE}
\end{equation}
For $d = 4$ (i.e.\ $\mathcal{N} = 4$ SYM), this yields
$S_{EE} = \sqrt{\lambda}/3$.

The capacity of entanglement $C_E = -\partial_n\widetilde{S}_n|_{n=1}$~\eqref{eq:capacity} requires
the slope of the refined R\'enyi entropy,
\begin{equation}
\partial_n\widetilde{S}_n\big|_{n=1}
= -\sqrt{\lambda}\bigl[2\zeta_h'(1) + \zeta_h''(1)\bigr]
= -\sqrt{\lambda}\;\frac{d-2}{(d-1)^3}\,.
\label{eq:mainResult}
\end{equation}
and is thus
\begin{equation}
C_E = \frac{\sqrt{\lambda}\,(d-2)}{(d-1)^3}\,,
\label{eq:CE_result}
\end{equation}
which for $d = 4$ gives $
C_E = \frac{2\sqrt{\lambda}}{27}$.

%
This is strictly positive for $d > 2$. By
Lemma~1 of Ref~\cite{Cao:2024nrx}, a nonvanishing capacity
of entanglement implies a non-flat entanglement spectrum,
which in turn establishes the presence of nonlocal magic
$\mathcal{M}^{(\rm NL)}$ in the Schwinger pair production. 
Note that $C_E$ depends only on the spacetime dimension~$d$ and the 
't~Hooft coupling~$\lambda$, and hence
reflects the intrinsic structure of the pair's quantum state in the 
gauge theory.

The physical picture is as follows. The worldsheet wormhole encodes the 
entanglement between the quark and antiquark and the total entanglement 
entropy is determined by the horizon area (position) at $n=1$. The 
structure of the entanglement spectrum, however, is encoded 
in the response of the horizon position to changes in $n$. 
This response is governed by the full $n$-dependent topological 
black hole family~\eqref{eq:fn}. For $d>2$, the topological black 
hole acquires a nontrivial ``charge'' that depends on $n$ and the geometry genuinely deforms under the replica parameter.

In $d=2$, the topological black hole is three-dimensional (the BTZ 
black hole), and the $n$-deformed geometry is locally equivalent 
to the undeformed one. $C_E$ and hence the antiflatness vanish. This does not exclude subleading 
$1/N$ or finite-$\lambda$ corrections which could yield a finite $C_E$ at subleading orders.

A possible field-theory interpretation of the vanishing of $C_E$ in $d=2$ is the
following. The bipartition considered in this work is the standard spherical region of the boundary CFT. CHM maps the reduced density
matrix $\rho_A$ to a thermal density matrix on
$S^1_\beta\times H^{d-1}$, with $\beta=2\pi n$. Accordingly, $\widetilde S_n$ is the thermal entropy in this hyperbolic-space, while $C_E$ is the corresponding heat capacity-like response.

For $d=2$ the causal diamond of the entangling region is conformally equivalent to a Rindler wedge with no transverse spatial
directions. The corresponding modular Hamiltonian is the conformal image of a boost generator, and the CHM thermal problem lives on
$S^1_\beta\times H^1$. The Schwinger-pair contribution then shifts the entanglement entropy, but its leading contribution to the CHM free energy is linear in the modular temperature. Consequently, its second derivative with respect to the modular temperature vanishes: the pair contributes to $S_{EE}$ but not to $C_E$. This should not be confused with the vacuum itself, whose whose capacity is nonzero.

For $d>2$, the same modular problem maps to a thermal theory on $S^1_\beta\times H^{d-1}$ with transverse hyperbolic directions. Hence, the Schwinger-pair contribution to the free energy is no longer linear in the modular temperature (which is the field-theory counterpart of the nonlinear replica dependence of
$\zeta_h(n)$) and $C_E>0$.

Note that the calculation implicitly contains the backreaction of the probe onto the spacetime which is necessary to compute any R\'enyi entropy to begin with. However, by employing the CHM map for the spherical entangling surface we can evade constructing the backreacted geometry explicitly since all necessary information is encoded in the free energy and can be extracted from there (see appendix~\ref{app:routeC_full} for details).

It is interesting to compare the ratio $C_E/S_{EE}$ in the 
for the probe correction with the vacuum result of~\cite{DeBoer:2018kvc,Nakaguchi:2016zqi}. 
For the vacuum of a $d$-dimensional CFT with holographic dual restricted to Einstein gravity and to a fixed regularization scheme, 
$C_E/S_{EE} =1$ for a spherical region. This result is tied to 
the conformal collider bounds~\cite{Hofman:2008ar,
Hofman:2016awc}. For the Wilson insertion, we find the coupling independent 
ratio of vacuum subtracted capacity to vacuum subtracted entanglement entropy
$\frac{C_E}{S_{EE}} 
= \frac{d-2}{(d-1)^2}$,
which only depends on the spacetime dimension. For $d=4$, 
$C_E/S_{EE} = 2/9$.

\section{Discussion}

We have shown that nonlocal magic is generated dynamically in 
Schwinger pair creation at strong coupling. The capacity of 
entanglement,
%
$C_E = \frac{\sqrt{\lambda}\,(d-2)}{(d-1)^3}\,,$
%
is strictly positive for $d>2$, establishing that the produced 
quark--antiquark pair carries genuinely nonlocal quantum correlations 
that go beyond entanglement. The color-singlet pair, despite its 
structural similarity to a Bell state, does carry nonlocal magic.

Notably, this result is independent of the acceleration $a = E/M$ 
and hence of the Unruh temperature $T_U = a/2\pi$. The nonlocal 
magic is therefore not a thermal artifact of the Rindler horizon, 
but reflects the intrinsic structure of the pair's quantum state as 
dictated by the gauge interaction for a given spacetime dimension.

The ratio $
C_E/S_{EE}
= (d-2)/(d-1)^2,
$
is independent of the 't Hooft coupling and reduces to $2/9$ for $d=4$.
It characterizes how far the entanglement spectrum deviates from 
flatness relative to its total weight.

The CHM map for a spherical entangling region produces a topological black hole whose equation of state has a curvature correction proportional to $(d-2)$. 
A probe string in this background has an action linear in $\zeta_h(n)$, so the
$n$-dependence of all R\'enyi quantities is governed entirely by $\zeta_h(n)$. The curvature correction renders $\zeta_h(n)$ nonlinear in
$1/n$, producing $C_E \neq 0$. For a planar black hole $\zeta_h(n)$ is linear in $1/n$ and $C_E = 0$. The capacity of entanglement is therefore controlled by the spatial curvature of the black hole horizon, which in the CHM construction is the curvature of $H^{d-1}$. This is only nontrivial for $d>2$.

A possible field theory interpretation is as follows. Under the CHM map, the reduced density matrix $\rho_A$
of a spherical region is mapped to a thermal density matrix on
$S^1_\beta\times H^{d-1}$, with $\beta=2\pi n$. In this representation,
$\widetilde S_n$ is the thermal entropy and $C_E$ is the corresponding
heat-capacity-like response to changing the modular temperature.

For $d=2$, the CHM thermal problem lives on $S^1_\beta\times H^1$. Equivalently, the modular Hamiltonian is
the conformal image of a boost generator. The Schwinger-pair contribution to the CHM free energy is linear in the modular temperature, and
therefore shifts the entanglement entropy without adding a heat-capacity-like response. Thus, at leading order, the pair contributes
to $S_{EE}$ but not to $C_E$. This does not mean that the full
two-dimensional interval spectrum is flat; in fact, the vacuum has nonzero capacity. For $d>2$, the CHM thermal problem is instead on
$S^1_\beta\times H^{d-1}$, and the pair contribution to the free energy has nonlinear modular temperature dependence, producing a positive capacity of
entanglement.

This observation has a broader significance. Recent 
work~\cite{Cao:2024nrx} has established that gravitational 
backreaction encodes nonlocal magic in holographic systems. Our 
result demonstrates that it 
can be recovered from the free energy in the probe limit.  The probe calculation should not be interpreted as neglecting
backreaction. The Wilson-line insertion changes the full gauge-theory state, and in the bulk this change is represented by
the first-order gravitational response of the geometry. The simplification
for spherical regions is that the CHM replica construction maps the calculation to the free energy (which is on shell). Thus constructing the gravitational response is
unnecessary, not the backreaction itself. This opens the 
door to studying nonlocal magic across the wide landscape of 
probe-brane systems in holography, including, for example defects, impurities, 
and flavor sectors, where entanglement has been extensively studied 
but complexity measures remain largely unexplored.

Several extensions merit investigation. In heavy-ion collisions, the electric fields are not constant but pulsed and it would be interesting to study the modification of the nonlocal magic during Sauter pulses~\cite{Grieninger:2023pyb,Florio:2021xvj}. It would be intriguing to understand the nonlocal magic during black hole evaporation and in the context of quantum extremal surfaces~\cite{Engelhardt:2014gca,Almheiri:2020cfm,Almheiri:2019qdq,Almheiri:2019psf,Almheiri:2019hni,Geng:2025byh,Geng:2025efs,Geng:2020qvw,Geng:2024xpj}. For example, it would be interesting to answer how nonlocal quantum correlations change across the 
Page curve in black hole evaporation. Third, one should extend our calculation to all values of the coupling~$\lambda$ via 
localization~\cite{Pestun:2007rz,Lewkowycz:2013laa,Pestun:2016zxk} which would require 
the connected stress-tensor correlator 
$\langle T_{\tau\tau}\,T_{\tau\tau}\rangle_W^{\mathrm{conn}}$ on 
$S^1\times H^3$ in the presence of the Polyakov loop. Moreover, the authors of~\cite{Amorosso:2024leg,Amorosso:2024glf,Amorosso:2026mdo} investigated the entanglement of the flux tube in Yang-Mills theories on the lattice using, among other observables, R\'enyi entropies and it would be interesting to understand the connection to our results.

Finally, the dynamical transition from zero to nonzero nonlocal 
magic in pair creation provides a concrete, analytically controlled 
example of complexity generation in a nonequilibrium process, which 
may be amenable to study via quantum simulation of lattice gauge 
theories.

\section*{Acknowledgements}
We thank Charles Cao and Andreas Karch for insightful discussions and comments on the draft.
This work is supported by a collaboration between the US DOE and other Agencies. This material is based upon work supported by the U.S. Department of Energy, Office of Science, National Quantum Information Science Research Centers, Quantum Systems Accelerator (Award No. DE-SCL0000121). Additional support is acknowledged from the U.S. Department of Energy, Office of Science, Office of Nuclear Physics, Inqubator for Quantum Simulation (IQuS) under Award Number DOE
(NP) Award DE-SC0020970 (S.G.). This work was also supported by the U.S. Department of Energy, Office of Science, Office of Nuclear Physics, Grant No. DE-FG02-97ER-41014 (UW Nuclear Theory, S.G.). S.G. was supported in part by a Feodor Lynen Research fellowship of the Alexander von Humboldt foundation. This work was also supported, in part, by the Department of Physics and the College of Arts and Sciences at the University of Washington.

\bibliography{refs}

\appendix
\section{Antiflatness as a diagnostic of nonlocal magic}\label{app:af}

In this section we summarize the information-theoretic relation between 
the structure of the entanglement spectrum and nonlocal quantum correlations, 
following~\cite{Cao:2024nrx} closely, and explain how this structure is accessed 
holographically.
  
To probe the nonlocal magic, it is useful 
to consider the response of the Rényi entropies under variations of the replica 
parameter $n$,
\begin{equation}
S_n = \frac{1}{1-n}\log \mathrm{Tr}\,\rho_A^n.
\end{equation}
A convenient quantity that isolates this response is the refined Rényi entropy,
\begin{equation}
\widetilde S_n \equiv n^2 \partial_n\left(\frac{n-1}{n}S_n\right),
\label{eq:refinedRenyi}
\end{equation}
which is sensitive to the $n$-dependence of the spectrum rather than its overall size.

Expressed in terms of the eigenvalues $\{\lambda_k\}$ of $\rho_A$, the derivative 
of $\widetilde S_n$ takes the form~\cite{Cao:2024nrx}
\begin{equation}
\partial_n \widetilde S_n
=
- n \frac{\sum_{k< l}\lambda_k^n \lambda_l^n 
\log^2\!\left(\frac{\lambda_k}{\lambda_l}\right)}
{\left(\sum_j \lambda_j^n\right)^2},
\label{eq:antiflatness_general}
\end{equation}
which is non-positive and vanishes only when all eigenvalues are equal. 
This shows that $\partial_n \widetilde S_n$ provides a direct measure of 
deviations from a flat entanglement spectrum, which we refer to as 
\emph{antiflatness}.

At $n=1$, this quantity admits a particularly simple interpretation as the 
capacity of entanglement~\cite{DeBoer:2018kvc},
\begin{equation}
\left.\partial_n \widetilde S_n\right|_{n=1}
=
- C_E(\rho_A)
=
- \mathrm{Var}_\rho(\log \rho),
\end{equation}
i.e., the variance of the modular Hamiltonian spectrum. Unlike the entropy, 
which depends only on the mean value, $C_E$ measures fluctuations of the 
entanglement spectrum.

We use the information-theoretic criterion of
Ref~\cite{Cao:2024nrx}, which relates the antiflatness of the entanglement spectrum
to nonlocal magic for a bipartite pure state. Importantly, this connection is 
purely information-theoretic and does not rely on holography. In particular:

\begin{enumerate}
\item \textbf{Faithfulness:} The capacity of entanglement vanishes if and only 
if the entanglement spectrum is flat. This is precisely the condition under 
which the nonlocal magic vanishes,
\begin{equation}
C_E(\rho_A) = 0 
\quad \Longleftrightarrow \quad 
\mathcal{M}^{(NL)}(\psi_{AB}) = 0.
\end{equation}

\item \textbf{Lower bound:} The antiflatness 
$F(\rho_A) = \mathrm{Tr}(\rho_A^3) - [\mathrm{Tr}(\rho_A^2)]^2$ 
provides a lower bound on the nonlocal magic,
\begin{equation}
\frac{F(\rho_A)}{8} \;\leq\; \mathcal{M}^{(NL)}_{\mathrm{dist}}(\psi_{AB}).
\label{eq:magic_bound}
\end{equation}

\item \textbf{Approximate relation:} When higher moments of the spectrum are 
subleading, the nonlocal stabilizer Rényi entropy is approximately proportional 
to the slope of the refined Rényi entropy,
\begin{equation}
\mathcal{M}_2^{NL}(\psi_{AB}) 
\;\approx\; 
\kappa\, \left|\partial_n \widetilde S_n\right|_{n=1},
\label{eq:proportionality}
\end{equation}
where $\kappa$ depends on the detailed spectral shape. In this regime, the 
capacity of entanglement is related to the antiflatness via
\begin{equation}
C_E(\rho_A) \approx \frac{F(\rho_A)}{\mathrm{Pur}(\rho_A)^2}.
\label{eq:CE_F_relation}
\end{equation}
\end{enumerate}

This shows that the slope of the refined Rényi entropy provides 
a quantitative measure of the internal structure of entanglement. In particular, 
states with identical entanglement entropy can exhibit very different values 
of $\partial_n \widetilde S_n$, reflecting differences in how correlations are 
distributed across the spectrum. Hence, $\partial_n \widetilde S_n$ is a more 
sensitive probe of genuinely nonclassical correlations than the entropy alone.

\section{Holographic setup and probe action}
\label{app:setup_full}

\subsection{Casini--Huerta--Myers map and topological black holes}
\label{app:CHM_full}

To compute R\'enyi entropies for a spherical entangling region
of radius~$R_A$, we employ the conformal mapping of Casini,
Huerta, and Myers~\cite{Casini:2011kv}.  Since the domain of dependence of a
spherical region is conformally equivalent to a Rindler wedge, this may equivalently
be viewed as a half-space bipartition after a conformal transformation. Starting from
Euclidean $\mathrm{AdS}_{d+1}$ in Poincar\'e coordinates,
\begin{equation}
ds^2 = \frac{L^2}{r^2}\, dr^2
+ \frac{r^2}{L^2}\!\left(dt_E^2 + d\rho^2
+ \rho^2\, d\Omega_{d-2}^2\right),
\label{eq:Poincare_app}
\end{equation}
we introduce dimensionless coordinates $(\zeta, u, \tau)$
via~\cite{Emparan:1999gf,Casini:2011kv}
\begin{align}
r &= \frac{L^2}{R_A}\bigl(\zeta\cosh u
+ \sqrt{\zeta^2 - 1}\,\cos\tau\bigr)\,,
\nonumber\\[3pt]
t_E &= R_A\,\frac{\sqrt{\zeta^2 - 1}\,\sin\tau}
{\zeta\cosh u + \sqrt{\zeta^2 - 1}\,\cos\tau}\,,
\label{eq:CHM_map_app}\\[3pt]
\rho &= R_A\,\frac{\zeta\sinh u}
{\zeta\cosh u + \sqrt{\zeta^2 - 1}\,\cos\tau}\,,
\nonumber
\end{align}
with $\zeta \in [1,\infty)$, $u \in [0,\infty)$, and
$\tau \sim \tau + 2\pi$. The metric becomes
\begin{equation}
ds^2 = L^2\!\left[\frac{d\zeta^2}{f(\zeta)}
+ f(\zeta)\, d\tau^2
+ \zeta^2\, du^2
+ \zeta^2 \sinh^2\!u\; d\Omega_{d-2}^2\right],
\label{eq:hyp_metric_app}
\end{equation}
with $f(\zeta) = \zeta^2 - 1$.

\subsection{R\'enyi entropies from the probe action}
\label{app:renyi_full}

Following the approach of Lewkowycz and 
Maldacena~\cite{Lewkowycz:2013nqa} in probe-brane 
language~\cite{Karch:2014ufa,Chalabi:2020tlw}, the probe 
contribution to the entanglement entropy relies on two inputs: the replica symmetry and the vanishing of the cross-term between bulk 
and probe actions at each~$n$ (see Appendix~\ref{app:routeC_full} 
for a detailed discussion of this point in the context of the 
backreaction calculation).

The probe string sits at a fixed point on $\mathbb{H}^{d-1}$ 
(for example $u = 0$, the quark location) and extends radially in~$\zeta$ 
from the horizon~$\zeta_h(n)$ to the Dirichlet-brane. In the CHM 
coordinates, the brane at Poincar\'e position $z = z_M$ maps to a 
position $\zeta_{\rm brane}$ that depends on $z_M$ and $R_A$ but is 
independent of~$n$: the brane sits at a fixed locus in the 
physical spacetime, and the $n$-dependence enters only through the 
bulk geometry below it.

The induced worldsheet metric for the string spanning $(\tau, \zeta)$ 
in the $n$-th background is
\begin{equation}
ds^2_{\rm ws} = L^2\!\left[\frac{d\zeta^2}{f_n(\zeta)} 
+ f_n(\zeta)\,d\tau^2\right],
\end{equation}
with $\sqrt{\det\gamma_{\rm ws}} = L^2$. The Nambu--Goto action 
over $\tau \in [0, 2\pi)$ is therefore
\begin{equation}
\widehat{I}(n) = \sqrt{\lambda}
\int_{\zeta_h(n)}^{\zeta_{\rm brane}} d\zeta 
= \sqrt{\lambda}\bigl(\zeta_{\rm brane} - \zeta_h(n)\bigr)\,.
\label{eq:Ihat_finite}
\end{equation}
The string has finite length, running from the horizon to the brane, 
and the action is manifestly finite. Since $\zeta_{\rm brane}$ is 
$n$-independent, all R\'enyi quantities depend only on $\zeta_h(n)$. 
We may therefore write
\begin{equation}
\widehat{I}(n) = -\sqrt{\lambda}\;\zeta_h(n) + \text{const}\,,
\label{eq:Ihat_result}
\end{equation}
where the constant $\sqrt{\lambda}\,\zeta_{\rm brane}$ drops out of 
every $n$-derivative and plays no role in what follows.

For the fundamental string, boundary contributions at the horizon 
vanish to all orders in 
$n$-derivatives~\cite{Kumar:2017vjv,Chalabi:2020tlw}. The refined 
R\'enyi entropy~\cite{Dong:2016fnf} is
\begin{equation}
\widetilde{S}_n = n^2\, \partial_n \widehat{I}(n) 
= -\sqrt{\lambda}\;n^2\,\zeta_h'(n)\,,
\label{eq:refined_renyi_app}
\end{equation}
and the capacity of entanglement is
\begin{equation}
C_E = -\partial_n \widetilde{S}_n \big|_{n=1}\,.
\label{eq:capacity_app}
\end{equation}
By the information-theoretic result of Ref~\cite{Cao:2024nrx}, 
$C_E > 0$ is both necessary and sufficient for nonlocal magic.

\section{Connecting the probe action to backreaction}
\label{app:equivalence}

In this appendix we establish that our calculation in the main text implicitly does come from the change in geometry due to the probe pair (route B). However, it becomes apparent that the change in geometry does in fact not have to be computed explicitly.

\subsection{Route A: Horizon area of topological black hole}
\label{app:routeA_full}

This is the direct computation from the replica construction, 
by carrying out the horizon derivatives
\begin{equation}
C_E = \sqrt{\lambda}\bigl[2\zeta_h'(1) + \zeta_h''(1)\bigr] 
= \frac{\sqrt{\lambda}\,(d{-}2)}{(d{-}1)^3}\,.
\end{equation}
The entanglement entropy $S_{\rm EE} = -\sqrt{\lambda}\,\zeta_h'(1) 
= \sqrt{\lambda}/(d{-}1)$ depends on the slope of the 
horizon trajectory $\zeta_h(n)$; the capacity $C_E$ depends on its 
curvature.

\subsection{Thermal interpretation}
\label{app:routeB_full}

The CHM map sends $\rho_A$ to a thermal state on 
$S^1_\beta \times \mathbb{H}^{d-1}$ at $\beta = 2\pi n$. The probe 
partition function is $Z = e^{n\widehat{I}(n)}$ (recall 
$\widehat{I} < 0$ in our conventions). The free energy is
\begin{equation}
F = -T\ln Z = -\frac{\widehat{I}(n)}{2\pi}\,.
\label{eq:free_energy}
\end{equation}
The thermal entropy is $S = -\partial F/\partial T$. Using 
$T = 1/(2\pi n)$:
\begin{equation}
S_{\rm thermal}(n) 
= -\frac{dF/dn}{dT/dn} 
= n^2\,\widehat{I}'(n) 
= \widetilde{S}_n\,.
\label{eq:key_identity}
\end{equation}
The refined R\'enyi entropy is identically the thermal entropy 
at $T = 1/(2\pi n)$. This holds as a function of~$n$ (including the well known case of~$n=1$ computing the entanglement entropy).

The heat capacity at fixed spatial geometry is 
$C_V = T\,dS/dT = -n\,\partial_n S$, giving at $n=1$:
\begin{equation}
C_V\big|_{T=1/(2\pi)} 
= -\partial_n\widetilde{S}_n\big|_{n=1} = C_E\,.
\label{eq:CV_equals_CE}
\end{equation}
Since $C_V = -T\,\partial_T^2 F$ and the free 
energy~\eqref{eq:free_energy} is obtained from the probe action on 
the fixed $n$-th background, $C_E$ is a free-energy-level 
quantity that does not require the backreacted metric.
In particular $C_E$ is proportional ($\sim\beta^2$) to the variance of
the modular Hamiltonian. 

\subsection{Route B: Backreaction}
\label{app:routeC_full}

We now show that route A implicitly incorporated backreaction by deriving the same result follows from the backreaction of 
the probe string on the bulk geometry, following the approach of 
Ref~\cite{Karch:2014ufa}. This makes it explicit how $C_E$ 
encodes the gravitational response to the replica 
deformation.

The probe string in the $n$-th topological black hole background 
backreacts on the ambient metric by the stress-energy tensor $T^{\mu\nu}_{\rm probe}(x)$
localized at $u = 0$ on $\mathbb{H}^{d-1}$ and extending from 
$\zeta_h(n)$ to $\zeta_{\rm brane}$ along the radial direction. 
In the probe expansion, the leading backreaction is
\begin{equation}
[\delta g(n)]_{\mu\nu}(x) 
= 8\pi G_N\! \int \!d^{d+1}x'\sqrt{g}\;
G^{(n)}_{\mu\nu\rho\sigma}(x,x')\;
T^{\rho\sigma}_{\rm probe}(x')\,,
\label{eq:delta_g}
\end{equation}
where $G^{(n)}_{\mu\nu\rho\sigma}$ is the graviton propagator 
in the $n$-th topological black hole 
background~\cite{DHoker:1999jke}. Both the propagator and the 
source depend on~$n$: the propagator through the background 
geometry, and the source through the lower integration limit 
$\zeta_h(n)$.

At each~$n$, the fixed-point set of the $\mathbb{Z}_n$ replica 
symmetry is the surface $\zeta = \zeta_h(n)$. The probe 
backreaction~\eqref{eq:delta_g} perturbs the area of this surface. 
Since the surface extremizes the area functional in the unperturbed 
background, the embedding perturbation drops out at leading 
order~\cite{Nozaki:2013vta,Chang:2013mca,Karch:2014ufa}, and the area change is 
given by the double integral~\cite{Chang:2013mca}:
\begin{equation}
\mathcal{K}(n) = \pi T_0 \int\!\!\int 
T^{\mu\nu}_{\rm min}(w)\;
G^{(n)}_{\mu\nu\rho\sigma}(w,z)\;
T^{\rho\sigma}_{\rm probe}(z)\,,
\label{eq:karch}
\end{equation}
where $T^{\mu\nu}_{\rm min}$ is the stress tensor of the extremal 
surface at $\zeta_h(n)$ (proportional to its induced metric). This 
is the gravitational potential energy between the extremal surface 
and the probe, mediated by the graviton propagator in the $n$-th 
background.

The total on-shell action on the $n$-fold replica manifold is
\begin{equation}
I_{\rm total}(n) 
= I_{\rm bulk}\!\left[g_0(n) 
  + \tfrac{\sqrt{\lambda}}{N^2}\,g_1(n)\right]
+ \sqrt{\lambda}\;\widehat{I}\!\left[g_0(n) + \cdots\right].
\label{eq:Itotal}
\end{equation}
Expanding in the probe parameter 
$\varepsilon = \sqrt{\lambda}/N^2 \ll 1$:
\begin{align}
I_{\rm total}(n) 
= &I_{\rm bulk}[g_0(n)] 
  + \varepsilon\;\frac{\delta I_{\rm bulk}}{\delta g}
    \bigg|_{g_0(n)}\!\!g_1(n)
  \nonumber\\&+ \sqrt{\lambda}\;\widehat{I}[g_0(n)]
  + \mathcal{O}(\varepsilon^2)\,.
\label{eq:Itotal_expanded}
\end{align}
The key step is that in the special case of spherical entangling surfaces the $n$-th topological black hole $g_0(n)$ is 
an exact solution of the vacuum Einstein equations with negative cosmological 
constant for every value of~$n$ and thus vanishes on-shell.
In other words, the cross-term between the bulk action and the probe 
backreaction vanishes. The probe contribution to the 
generalized gravitational entropy is thus captured by 
$\widehat{I}(n)$ on the fixed background, proving the equivalence
\begin{equation}
\mathcal{K}(n) = \partial_n\widehat{I}(n)
\label{eq:K_equals_Iprime}
\end{equation}
between the integral~\eqref{eq:karch} and the 
$n$-derivative of the probe action~\eqref{eq:Ihat_result} at 
every~$n$.
\end{document}